# Comprehensive Evaluation and Insights into the Use of Large Language Models in the Automation of Behavior-Driven Development Acceptance Test Formulation

Shanthi Karpurapu[1], Sravanthy Myneni[2], Unnati Nettur[3], Likhit Sagar Gajja[4], Dave Burke[5], Tom Stiehm[6], AND Jeffery Payne[7]
Corresponding author: Shanthi Karpurapu (shanthi.karpurapu@gmail.com)

**ABSTRACT** Behavior-driven development (BDD) is an Agile testing methodology fostering collaboration among developers, QA analysts, and stakeholders. In this manuscript, we propose a novel approach to enhance BDD practices using large language models (LLMs) to automate acceptance test generation. Our study uses zero and few-shot prompts to evaluate LLMs such as GPT-3.5, GPT-4, Llama-2-13B, and PaLM-2. The paper presents a detailed methodology that includes the dataset, prompt techniques, LLMs, and the evaluation process. The results demonstrate that GPT-3.5 and GPT-4 generate error-free BDD acceptance tests with better performance. The few-shot prompt technique highlights its ability to provide higher accuracy by incorporating examples for in-context learning. Furthermore, the study examines syntax errors, validation accuracy, and comparative analysis of LLMs, revealing their effectiveness in enhancing BDD practices. However, our study acknowledges that there are limitations to the proposed approach. We emphasize that this approach can support collaborative BDD processes and create opportunities for future research into automated BDD acceptance test generation using LLMs.

## I. INTRODUCTION

BDD is a widely adopted, test-first Agile testing practice for specifying system behavior through tests. It promotes collaboration among software developers, quality assurance analysts, and business stakeholders [1] [5] [35]. BDD utilizes the Gherkin language [2] [33], a semi-structured natural language, to express the acceptance criteria of user stories in human-readable acceptance tests written as scenarios framed by the Given, When, Then keywords [3].

A simple scenario is provided in Table 1 to illustrate the use of the Gherkin language. In Table 1, The Given step establishes the initial condition, ensuring the calculator is ready for use. The When step represents the action – entering the numbers 5 and 7 and choosing addition. Finally, the Then step defines the expected outcome – the calculator displays the result as "12". This Gherkin structure enhances clarity and enables non-technical professionals to understand scenarios.

**TABLE 1: BDD Scenario example**

| |
|---|
| Scenario: Performing Addition |
|   Given I have opened the calculator application |
|   When I enter "5" into the calculator |
|   And I add "7" |
|   Then the result should be "12" |

Despite its effectiveness in translating software specifications into behavior for validation, BDD experiences challenges in manually crafting acceptance tests, leading to a notable impact on efficiency and accuracy [4]. Ensuring high-quality BDD acceptance tests is crucial for successful implementation [34]. The term "high-quality BDD acceptance tests" denotes adherence to Gherkin rules and best practices outlined in the methodology section [6]. In addition, significant experience is required to craft good quality BDD acceptance tests.

In this manuscript we describe an approach that optimizes BDD practices using LLMs, aiming to automate acceptance test generation, reduce manual effort, and enhance productivity. The suggested approach can also aid in standardizing the BDD acceptance test formulation in Agile teams. Our study explores the automation of BDD acceptance test possibilities using LLMs such as GPT-3.5, GPT-4, Llama-2-13B, and PaLM-2. Additionally, it delves into the significance of prompt engineering for evaluating and selecting optimal prompts to ensure the generated scenarios adhere to best practices and avoid Gherkin syntax issues.

## II. RELATED WORKS

This section provides an overview of our literature review on BDD acceptance test creation automation. Despite our exhaustive search, we found no prior research in this field. Therefore, we broadened our literature review to explore the automation of conventional test case generation methods. The term 'conventional test case generation method' refers to test cases written with preconditions, test steps, expected results, and actual results. We have reviewed approaches that used conventional based methods.

Mathur et al. [7] proposed a method for generating test cases from conversation text. The context and topic are extracted from the conversation text using T5 and GPT-3 models. This approach eliminates the need for manual test case creation and enables the testing of complex systems. However, the method may need to enhance its performance when generating numerous test cases and exhibiting accuracy issues in the output.

Zhuang et al. [8] proposed a technique in 2018 for generating short test cases using an ensemble model. This method uses multiple rankers and a generation-based approach to produce the final response, which is selected based on the ensemble module ranking provided. However, Interpreting the output of an ensemble model can be challenging due to its complexity. Additionally, creating an ensemble model can be challenging and time-consuming, which may result in lower prediction accuracy compared to using a single model.

In a different study conducted by Lafi et al. [9] proposed an approach to generate test cases automatically from requirement specifications. This approach has a sequence of steps: extracting the required information from the use case description obtained from the use case diagram, then developing the control flow graph and NLP table, and generating test paths as the next step. The test paths and NLP table generate test cases, but there is a need to explore the applicability of the proposed approach to various case studies. This approach does not address validating results obtained using different test coverage criteria.

Guo et al. [10] introduced a framework for generating test cases based on the generative adversarial network (GAN). This study has shown that the WGAN-GP model effectively improves test coverage. The results indicate that unit testing has better test coverage than integration testing. However, the study highlights the potential for enhancing integration tests in the future.

Utting et al. [11] developed an approach for efficient test identification and generation from the customer and test execution traces, particularly in web services and API testing. This approach utilizes clustering algorithms to improve testing procedures. This study only presents preliminary results and has yet to determine the method's robustness.

Our proposed approach uses LLMs like GPT-3.5, GPT-4, or PaLM-2 with optimized prompt design deviating from machine learning-based conventional test case generation methods [12] – [15] mentioned in the literature [12]. In the conventional approach, teams typically write test cases in the later stages of software development. Our approach is developed for BDD based agile methodology [40] and enables early testing and validation. Therefore, teams formulate BDD acceptance tests for user stories before progressing from the sprint backlog to the 'in progress' state, defining all specifications upfront in a testable format. Additionally, using LLMs in the proposed approach has the advantage of being simple, easily accessible to everyone, and user-friendly. Adopting this proposed approach is relatively straightforward for Agile teams, provided the organization's policies allow the use of LLMs.

## III. METHODOLOGY

Our proposed methodology aims to comprehensively evaluate different LLMs and prompt design techniques to gauge their effectiveness in generating BDD acceptance tests. Figure 1 illustrates the procedural steps for automating BDD acceptance tests and validation.

We developed separate programs for each LLM and executed each program twice to evaluate both few-shot and zero-shot prompts. The implementation is written in Python, tailored individually for each model, and executed within the Google Colab framework [17]. The code developed for all the models is available in the GitHub repository [16].

The initial step involves reading user story descriptions from a CSV file. Each user story description, accompanied by the prompt (prompt is the task instructions to the model) and model parameters (temperature, top_p, max_tokens etc.), are sent in the request body to the model API endpoint. Iterating through all user stories, we store the model's BDD acceptance tests (scenarios) response as a .feature file.

After the feature file generation phase, the generated feature files undergo validation using a tool that validates Gherkin syntax, capturing syntax errors for further analysis. We comprehensively assessed LLMs and prompt design techniques for this proposed methodology that integrates automated generation with rigorous validation.

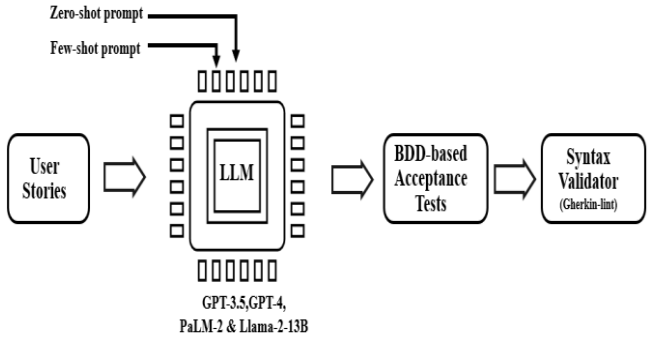

**FIGURE 1.** Behavior driven development acceptance tests generation and validation process.

The following details outline the essential elements of our proposed methodology.

### A. DATASET
For this study, we curated a diverse set of approximately 50 real-world user stories drawn from various sources. The data sources include user story data from Mendeley [18] and a blog post [19] that details user stories across different domains. This compilation ensures a comprehensive representation of user stories for robust examination in our study.

### B. PROMPT TECHNIQUES
The prompt technique involves creating clear, concise instructions or queries that guide LLMs in task execution. These prompts are essential because they enhance model comprehension and performance by providing explicit guidance, reducing ambiguity, and ensuring accurate responses. Wei et al. (2022) suggested the effectiveness of instruction tuning, which refines models through fine-tuning based on datasets described with clear instructions, thereby enhancing zero-shot learning capabilities [20] [36] [37].

In the context of few-shot learning, prompts are crucial. Despite the impressive zero-shot capabilities of LLMs, they may face challenges with complex tasks. Therefore, few-shot prompting [39] involves incorporating examples within the prompt. This prompt technique enables in-context learning by guiding the model to better performance, conditioning it for subsequent examples with minimal instances.

Adhering to best practices in generating BDD acceptance tests can fully unlock the value of LLM. We evaluated zero and few-shot prompt techniques [20] - [24] to determine which prompts better align with best practices. Best practices adoption may vary based on the software development practices followed by the agile teams. Therefore, as part of this study, we considered a subset of common best practices for evaluation, focusing solely on evaluating if the model can give an output based on the best practice instructions given in the prompt.

Table 2 and 3 details the prompt techniques considered in this study. The best practices are included in both the prompts as instructions from 1 to 6, as shown in Table 2 and 3. It is worth noting that zero and few-shot prompts have the same instructions, and the few-shot prompt includes additional BDD scenario examples for the model to follow while producing the response.

**TABLE 2: Zero-shot prompt technique instructions.**

| |
|---|
| Generate a feature file with 5 Gherkin Scenarios for {user_story} by following below {instructions}.<br>{instructions}=<br>1. Start the feature file with the 'Feature:' keyword.<br>2. Provide a descriptive feature name to specify the context of the scenarios.<br>3. Include steps in the Background if they are repeated at the beginning of all scenarios in a feature.<br>4. The background step is executed before every scenario.<br>5. Use tags as annotations to group and organize scenarios and features.<br>6. Tags are written with the '@' symbol followed by a significant text.<br>{user_story} = "As a user, I need a simple calculator for quick and accurate basic operations." |

**TABLE 3: Few--shot prompt technique instructions.**

| |
|---|
| User role:<br>Same instructions from Zero-shot are given along with the examples as written below<br>Assistant role:<br>Feature: Basic Calculator Operations<br>  As a user, I need a simple calculator for quick and accurate basic operations.<br>    Background:<br>      Given I have opened the calculator application<br>    @basicoperations<br>    Scenario: Performing Addition<br>      When I enter "5" into the calculator<br>      And I add "7"<br>      Then the result should be "12"<br>    @basicoperations<br>    Scenario: Performing Subtraction<br>    When I enter "10" into the calculator<br>      And I subtract "3"<br>      Then the result should be "7"<br>  # Continued. …….. |

### C. LARGE LANGUAGE MODELS

We listed the LLMs considered for evaluation, as detailed in Table 4. The selected models include:

**TABLE 4: Large Langue Model Details.**

| Model | Details |
|---|---|
| GPT-3.5-Turbo | This autoregressive LLM utilizes the Reinforcement Learning from Human Feedback (RLHF) mechanism. As the initial backbone model for ChatGPT, GPT-3.5 [38] demonstrates remarkable zero-shot performance across diverse tasks [25]. Our study employs the gpt-3.5-turbo-0613 model from OpenAI, known for its maximum context length of 4K tokens. |
| GPT-4-Preview | The most recent addition to OpenAI's LLMs, GPT-4, stands out as the first multimodal model in the GPT series [26]. Recognized for its enhanced instruction-following capabilities compared to GPT-3.5, GPT-4 is deemed more reliable. In our study, we utilize the GPT-4 version. |
| PaLM-2 | Developed by Google [27], PaLM-2 is an LLM that employs a mixture of objective techniques It demonstrates substantial performance improvements over the original PaLM model [28]. In our study, we utilize the chat-bison-001 model. |
| Llama-2 – 13B | Developed by Meta [29], Llama-2 stands out as an open-source LLM. A notable advantage of Llama-2, distinguishing it from the previously mentioned LLMs, is its open-source nature, making it accessible for research and commercial purposes. Our study utilizes the Chat versions of the Llama-2-13B model, sourced from HuggingFace [30]. |

### D. EVALUATION

Manual validation of feature files generated from LLM can be resource-intensive, demanding significant time and effort. In addition, the human-driven nature of the process introduces the potential for biases and errors, emphasizing the need for rigorous checks and collaboration to mitigate these challenges.

In evaluating the quality of the generated feature files, we employed the Gherkin-lint tool [31] [32] to overcome manual validation challenges, a robust instrument designed to detect syntax violations. This tool meticulously evaluates adherence to predefined style guidelines tailored for the Gherkin language. This tool comprehensively evaluates and examines the syntactical correctness of the feature files and ensures alignment with established conventions and best practices in Gherkin-based specifications. This approach offers a detailed assessment of BDD acceptance tests and enhances the robustness and comprehensiveness of our evaluation process.

## IV. RESULTS

As the methodology section outlines, we generate BDD acceptance tests for all user stories, considering various LLMs and prompts. The subsequent sections present the findings from the conducted experiments. Visual representations have been employed to convey critical findings and offer comprehensive insights into the experiments.

### A. VALIDATION ACCURACY

As shown in equation (1), we calculated accuracy by dividing the number of BDD feature files without syntax issues by the total number of generated BDD feature files. When employing the few-shot prompt technique, we observe a high success rate in generating BDD feature files without syntax issues. GPT-3.5 and GPT-4 models exhibit higher accuracy with the few-shot prompt technique than other models, such as Llama-2-13B and PaLM-2 (Figure 2). Llama-2-13B achieves the second-highest accuracy with the few-shot prompt technique (Figure 2).

$$Validation\ Accuracy = \frac{No.of\ Feature\ Files\ w/o\ Syntax\ Errors}{Total\ No.of\ Feature\ Files} \quad (1)$$

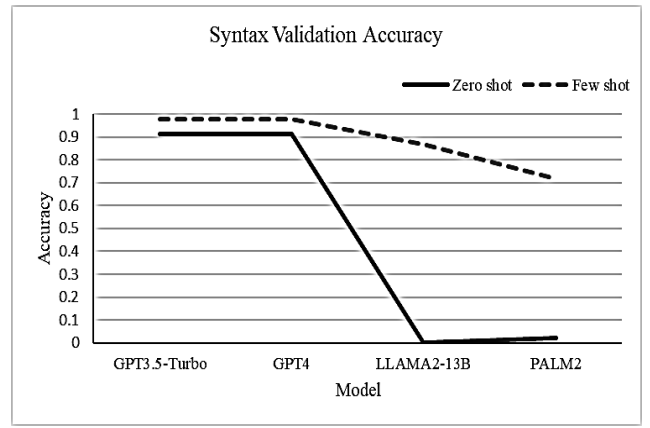

**FIGURE 2.** BDD-based feature syntax validation accuracies across all user stories with various prompt techniques and LLMs.

### B. SYNTAX ERRORS

The various syntax errors reported by the Gherkin-lint tool are categorized as shown in Table 5. The aggregation of syntax errors across all models reveals that the highest incidence of errors occurs when employing the zero-shot technique (Figure 3). Notably, the zero-shot approach contributes to 89% of the total syntax errors (Figure 3).

**TABLE 5: Syntax error type details**

| Error Type | Error Details |
|---|---|
| gherkin-keywords-not-in-logical-order | When writing scenario steps in some BDD feature files generated by LLM, the tool reports an error if the specified order of scenario step keywords (Given, When, Then) is not followed—for instance, a sequence like "Given Then When" is considered incorrect. |
| gherkin-keyword-not-present-in-step | The tool reports this error when scenario steps do not commence with the keywords Given, When, Then, And, or But. For instance, inappropriately, tags are sometimes written at the end of the file, such as @regression |
| | @download-files @experiments @attachments, violating the correct structure where tags should appear before the scenario description. |
| missing-tags | The tool reports this error in the BDD feature file when scenarios lack tags. While reporting the lack of tags error is optional in the tool settings, the configuration dictates treating missing tags as an error, as the prompt technique involves generating tags as one of the instructions. |
| restricted-patterns-present | The presence of unnecessary special characters in the generated BDD feature files triggers the reporting of this error. Examples include: Feature description: "-----------------------------------" (matches a restricted pattern) Background description: "------------" (matches restricted pattern) Scenario description: "-------------------------" |

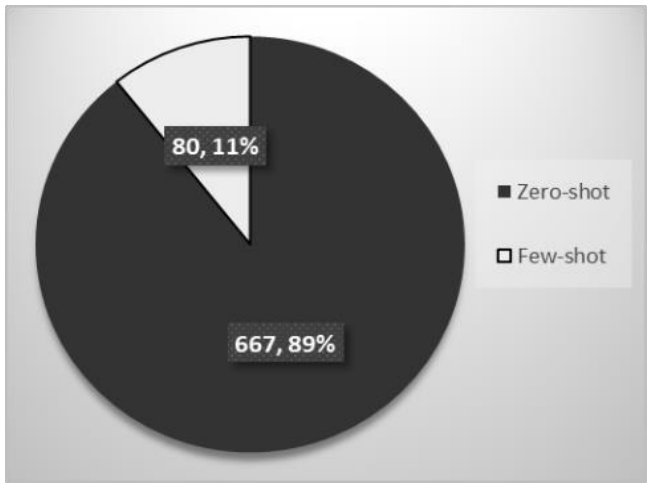

**FIGURE 3.** Total syntax errors for zero and few-shot techniques.

The compilation of syntax errors by error type across all models indicates that the predominant contributors to errors in the zero-shot prompt technique are the "Restricted-Patterns-Present" and "Gherkin-Keyword-Not-Present" error types. In the few-shot prompt technique, the most prominent error types are "Restricted-Patterns-Present" and "Missing-Tags (Figure 4).

The examination of model performance concerning syntax errors shown in Figure 5 is a vital aspect. GPT-3.5 and GPT-4 exhibit better performance by showcasing the fewest errors in the few-shot technique. Additionally, the number of syntax issues observed in the zero-shot technique for GPT-3.5 and GPT-4 is comparable but slightly higher than that in the few-shot technique. GPT-3.5 and GPT-4 are the most effective models, with fewer syntax errors in generating BDD feature files as shown in Figure 5. Conversely, Llama-2-13B contributes the highest number of errors in the zero-shot technique, while PaLM-2 does so in the few-shot technique.

**FIGURE 4.** Syntax error type distribution.

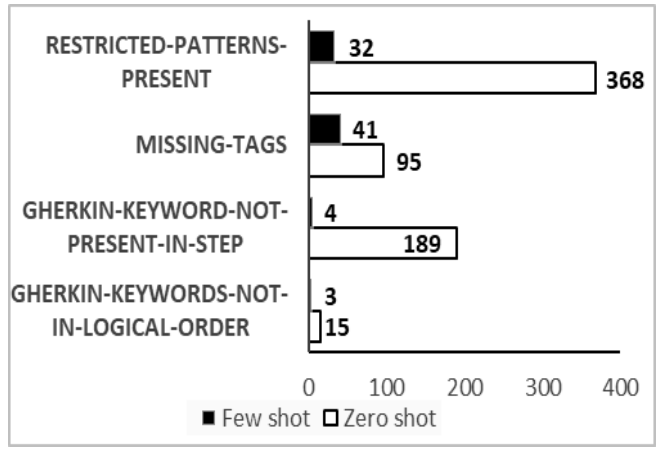

**FIGURE 5.** No. of syntax errors for various LLMs.

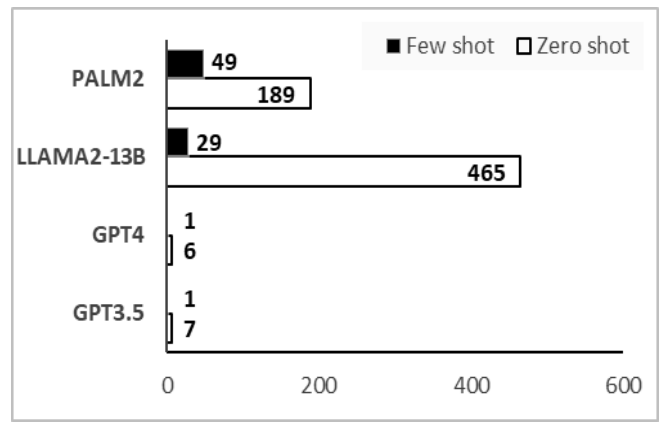

This comparative analysis evaluates GPT-3.5, GPT-4, Llama-2-13B, and PaLM-2 language models based on their performance in handling Gherkin syntax errors, including gherkin-keywords-not-in-logical-order, gherkin-keyword-not-present-in-step, missing-tags, and restricted-patterns-present. As shown in Table 6 & 7, GPT-3.5 and GPT-4 displayed limited occurrences of errors in zero-shot and few-shot evaluations. Llama-2-13B exhibited challenges during zero-shot evaluations, particularly in gherkin-keyword-not-present-in-step (130 instances) and restricted-patterns-present (335 instances). PaLM-2 demonstrated a spectrum of errors, with many instances involving missing-tags as (95 occurrences). Notably, few-shot evaluations across all models revealed an overall lower number of errors compared to zero-shot evaluations.

**TABLE 6:** Syntax error distribution for zero-shot.

| Error Type | GPT-3.5 | GPT-4 | PaLM-2 | Llama-2-13B |
|---|---|---|---|---|
| gherkin-keywords-not-in-logical-order | 6 | 5 | 4 | 0 |
| gherkin-keyword-not-present-in-step | 1 | 0 | 58 | 130 |
| restricted-patterns-present | 0 | 1 | 32 | 335 |
| missing-tags | 0 | 0 | 95 | 0 |

TABLE 7: Syntax error distribution for few-shot.

| Error Type | GPT-3.5 | GPT-4 | PaLM-2 | Llama-2-13B |
|---|---|---|---|---|
| gherkin-keywords-not-in-logical-order | 1 | 1 | 1 | 0 |
| gherkin-keyword-not-present-in-step | 0 | 0 | 3 | 1 |
| restricted-patterns-present | 0 | 0 | 4 | 28 |
| missing-tags | 0 | 0 | 41 | 0 |

Although GPT-3.5 and GPT-4 models performed better with the few-shot prompt technique, our proposed approach had certain limitations within the current research scope. We have manually validated some of the BDD acceptance tests and confirmed the validity of those tests in terms of their applicability to user stories. In our present research scope, we haven't explored the realms of test coverage and the validity of generated tests in terms of their applicability to user stories using any formal validation process. However, we recognize these areas as valuable aspects for future research. The foundational findings from our research should instill confidence and pave the way for a more in-depth examination of test coverage and validity in subsequent research in this field. Our primary aim was to validate the readiness of tests produced by the LLM model, ensuring they are free from syntax issues. Also, we employed a single syntax validation tool for validation. We assumed that the outcomes would likely align across other syntax validator tools.

Furthermore, in our current study, we couldn't utilize any real-time projects due to the lack of dataset, which can boost our confidence if explored with real-time project data. The BDD methodology emphasizes collaboration among stakeholders and developers for crafting acceptance tests. The suggested method should facilitate this collaborative process, signifying its role as a supportive tool. It implies that the approach should assist in the collaborative process and not be utilized in isolation, emphasizing the need for discussions on formulating acceptance tests from the proposed approach.

## V. CONCLUSION

Utilizing OpenAI's GPT-3.5 and GPT-4 models and the few-shot prompt technique to create BDD acceptance tests from user stories displays potential for improving productivity in agile software development and implementing standardized BDD practices. Incorporating the few-shot technique, involving instruction tuning with examples has demonstrated improved performance compared to the zero-shot approach. This advancement proves particularly beneficial in addressing complex natural language problems, such as the creation of BDD acceptance tests. The potential success of this approach could inspire further research into utilizing LLMs for automated BDD acceptance test generation to achieve enhanced test coverage.